\documentclass[pra,twocolumn, aps,letterpaper, preprintnumbers,superscriptaddress]{revtex4-1}
\usepackage{hyperref}
\usepackage{verbatim}
\usepackage{amsmath}
\usepackage{latexsym}
\usepackage{revsymb}
\usepackage{yfonts}
\usepackage{ifthen}

\usepackage{natbib}
\usepackage{amsfonts}
\usepackage{amsmath}
\usepackage{amssymb}
\usepackage{amsthm}
\usepackage{graphicx}
\usepackage{bm}
\usepackage{bbm}
\usepackage{epsfig,color,amssymb}
\usepackage{subfigure}
\usepackage{amsfonts}
\usepackage{amscd}
\usepackage{amsmath}
\usepackage{multirow}
\usepackage{chemarrow}
\usepackage{dcolumn}
\usepackage{bm}
\usepackage{graphicx}
\usepackage{enumerate}
\usepackage{epsfig}
\usepackage{subfigure}
\usepackage{xcolor}
\usepackage{multirow}
\usepackage{ulem}
\usepackage{braket}
\usepackage{comment}
\usepackage{enumitem}
\usepackage{amsthm}
\renewcommand{\emph}[1]{\textit{#1}}

\begin{document}

\title{Finite-key analysis for twin-field quantum key distribution with composable security}

\author{Hua-Lei Yin}\email{hlyin@nju.edu.cn}
\author{Zeng-Bing Chen}\email{zbchen@nju.edu.cn}
\affiliation{National Laboratory of Solid State Microstructures and School of Physics, Nanjing University, Nanjing 210093, China}

\begin{abstract}
Long-distance quantum key distribution (QKD) has long time seriously relied on trusted relay or quantum repeater, which either has security threat or is far from practical implementation.
Recently, a solution called twin-field (TF) QKD and its variants have been proposed to overcome this challenge. However, most security proofs are complicated, a majority of which could only ensure security against collective attacks. Until now, the full and simple security proof can only be provided with asymptotic resource assumption.
Here, we provide a composable finite-key analysis for coherent-state-based TF-QKD with rigorous security proof against general attacks.
Furthermore, we develop the optimal statistical fluctuation analysis method to significantly improve secret key rate in high-loss regime.
The results show that coherent-state-based TF-QKD is practical and feasible, with the potential to apply over nearly one thousand kilometers.
\end{abstract}

\maketitle

\section*{Introduction}
Classical encryption communication plays a central role in network security, which, however, faces increasingly serious security threats with quantum computation~\cite{fedorov2018quantum}. Quantum key distribution (QKD)~\cite{bennett1984quantum,Ekert:1991:Quantum} promises information-theoretically secure encryption communication with the laws of quantum mechanics.
However, in practice, there are two important problems severely restrict QKD implementations.
One is the rate-distance limit of QKD~\cite{takeoka2014fundamental}, which means that the secret key rate is linear scaling with channel transmittance and bounded by the secret-key capacity of quantum channel~\cite{takeoka2014fundamental,pirandola2017fundamental}.
It is believed that the limit of transmission distance is approximately 500 km ultralow-loss fibre~\cite{gisin2015far}.
The other is the quantum hacking attacks or, more precisely, the side-channel attacks on detection~\cite{lo2014secure}. In the security proof of typical QKD, one requires that the detection probability of signal is basis-independent. However, it is very easy to be broken without being detected, for example, by the detector blinding attack~\cite{lydersen2010hacking}. The big gap between experimental realizations and theoretical models on the measurement devices is often exploited by eavesdroppers to successfully steal the key.

To circumvent the rate-distance limit, the trusted relay~\cite{Liao:2018:Satellite} or quantum repeater~\cite{Sangouard:2011:Quantum} schemes are proposed. However, the trusted relay significantly compromise the security while the quantum repeater techniques are far from practical implementation. To overcome the side-channel attacks on detection, the measurement-device-independent (MDI) QKD based on two-photon Bell state measurement~\cite{lo2012measurement} has been proposed and experimentally demonstrated over 404 km ultralow-loss fibre~\cite{Yin:2016:Measurement}. Unfortunately, the secret key rate of MDI-QKD is far below typical QKD in realistic implementations~\cite{Yin:2016:Measurement,Boaron:2018:Secure}.

Recently, a novel protocol known as twin-field (TF) QKD~\cite{lucamarini2018overcoming} has been introduced to simultaneously solve the above two problems by exploiting the single-photon interference in the untrusted relay, which provides a secret key rate proportional to the square-root of channel transmittance and is immune to any attack on measurement devices. Until now, several proof-of-principle experimental demonstration of TF-QKD have already been successfully performed~\cite{Minder:2019:Experimental,Wang:2019:Beating,Liu:2019:Experimental,Zhong:2019:Proof}, indicating that the techniques of TF-QKD are realizable.
The original TF-QKD is a remarkable breakthrough in the field of quantum communication even without unconditional security proof. To prove the security of TF-QKD, two types of variants are proposed ~\cite{ma2018phase,tamaki2018information,Wang2018Sending,yin2018practical,yu2019sending,curty:2018:simple,cui:2018:phase,Lin:2018:A,primaatmaja2019almost,Yin:2018:Twin}.
One is the single-photon-based TF-QKD~\cite{Wang2018Sending,yin2018practical,curty:2018:simple} first proposed in Ref.~\cite{Wang2018Sending} named as sending-or-not-sending protocol with a security proof against coherent attack. It is similar with the original TF-QKD using the single-photon component to extract secret key by implementing single-photon Bell state measurement~\cite{yin2018practical,curty:2018:simple}. Recently, we became aware that the single-photon based protocol and its key rate formula, Eq. (3) presented in Ref. [22] are actually same with the earlier protocol [21], sending-or-not-sending protocol proposed by Wang et al. We thank authors of Ref. [21] for pointing out this.
The other is the coherent-state-based TF-QKD~\cite{ma2018phase,tamaki2018information,curty:2018:simple,cui:2018:phase,Lin:2018:A,primaatmaja2019almost,Yin:2018:Twin}, or called phase-matching QKD, which directly exploits the coherent state to extract secret key by implementing entangled coherent state measurement~\cite{Yin:2018:Twin}.
However, so far, taking into account all finite-size effects in TF-QKD with rigorously composable security proof is still missing, which severely influences TF-QKD to become as practical and feasible as typical QKD~\cite{tomamichel2012tight,lim2014concise} and MDI-QKD~\cite{curty2014finite} with composable security under realistic conditions.

In this work, we provide a composable finite-key analysis for coherent-state-based TF-QKD with rigorous security proof against general attacks. We make three contributions to obtain the optimal secret key rate and show that the transmission distance can surpass 800 km fibre with the realistic technology.
First, we use the entropic uncertainty relation~\cite{tomamichel2011uncertainty} to prove the security of coherent-state-based TF-QKD in the finite-key regime. 
It is known to all that entropic uncertainty relation is well suited for the composable security proof against general attacks, which is rather direct and avoids various estimations~\cite{tomamichel2012tight,curty2014finite,lim2014concise}.
Second, we develop the tight and rigorous multiplicative Chernoff bound and its variant to deal with the difference between the observed value and the expected value, which closes the gap between the large-deviation Chernoff bound method~\cite{curty2014finite} and the not-sufficiently-rigorous Gaussian analysis.
Third, the tailored tail inequality for random sampling without replacement is the tightest, which further improves the secret key rate in the finite-key regime.

\begin{figure}
\centering
\includegraphics[width=8cm]{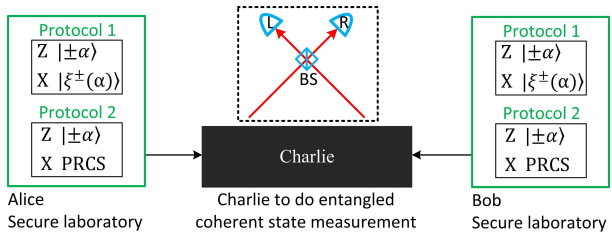}
\caption{The setup of coherent-state-based TF-QKD.
For Protocol 1 (2), Alice and Bob prepare coherent states $\ket{\pm\alpha}$ if choosing $\rm Z$ basis and cat states $\ket{\xi^{\pm}(\alpha)}$ (PRCS) if choosing $\rm X$ basis.
They send the prepared quantum signals through insecure channel to the untrusted Charlie, who is supposed to perform an entangled coherent state measurement. As an example, Charlie let the two received optical pulses interfere at a symmetric beam splitter (BS),
which has on each end a threshold single-photon detector. A click in the single-photon detector $L$ implies a projection into the entangled coherent state $\ket{\Phi^{-}}=1/\sqrt{N_{-}}(\ket{\alpha}\ket{\alpha}-\ket{-\alpha}\ket{-\alpha})$, while a click in single-photon detector $R$ indicates a projection into the entangled coherent state $\ket{\Psi^{-}}=1/\sqrt{N_{-}}(\ket{\alpha}\ket{-\alpha}-\ket{-\alpha}\ket{\alpha})$.
Details can be found in main text.
\label{fig1}}
\end{figure}

\section*{Results}

\noindent
\textbf{Security definition.}
Before introducing our protocol, we follow the discussion of the so-called universally composable framework~\cite{muller2009composability}. A general QKD protocol either outcomes a pair of key bit strings $\textbf{S}$ and $\hat{\textbf{S}}$ for Alice and Bob or aborts denoted by $\textbf{S}=\hat{\textbf{S}}=\perp$. The length of bit strings $\textbf{S}$ and $\hat{\textbf{S}}$ are both equal to $\ell$. In general, the QKD protocol is called secure if the
key bit strings satisfy two criteria, namely, the correctness and the secrecy criteria.

The correctness criterion is met if the key bit strings of Alice and Bob are identical, i.e., $\textbf{S}=\hat{\textbf{S}}$. However, the correctness criterion cannot be perfectly satisfied in experiment, which means that we may allow some negligible errors. Specifically, we say that a protocol is $\varepsilon_{\textrm{cor}}$-correct
if $\Pr[\textbf{S}\not=\hat{\textbf{S}}] \leq \varepsilon_{\textrm{cor}}$, i.e., the probability that Alice's and Bob's key bit strings are not identical does not exceed $\varepsilon_{\textrm{cor}}$.

Let system $\textbf{E}$ be the information of eavesdropper during the process of the QKD protocol, $\{\ket{s}\}_s$ be an orthonormal basis for Alice's system and $\rho_{\textbf{E}}^s$ be the state of the system $\textbf{E}$ given any fixed value $s$ of key bit string $\textbf{S}$.
In order to define secrecy, we should introduce a description of the correlation between the key bit string of Alice $\textbf{S}$ and eavesdropper, which can be given by the joint classical-quantum state $\rho_{\textbf{S}\textbf{E}}=\sum_{s}p_{s}\ket{s}\bra{s}\otimes \rho_{\textbf{E}}^s$. The secrecy criterion is met if the system $\textbf{E}$ completely has no correlation with the key bit string of Alice, i,e., $\rho_{\textbf{S}\textbf{E}}=U_{\textbf{S}} \otimes \rho_{\textbf{E}}$, where $U_{\textbf{S}}=\sum_s\frac{1}{|{\mathcal S}|}\ket{s}\bra{s}$ is the uniform mixture of all possible values of the key bit string $\textbf{S}$.
However, the secrecy criterion can still never be perfectly satisfied in experiment. We say that a protocol is $\varepsilon_{\rm sec}$-secret if the trace distance between the joint classical-quantum state $\rho_{\textbf{S}\textbf{E}}$ and the ideal case described by $U_{\textbf{S}} \otimes \rho_{\textbf{E}}$ is no more than $\Delta$, i.e.,
\begin{equation}
\begin{aligned}\label{eqll}
\frac{1}{2}\|\rho_{\rm \textbf{S}\textbf{E}}-U_{\textbf{S}}\otimes\rho_{\textbf{E}}\|_{1} \leq \Delta, \nonumber
\end{aligned}
\end{equation}
and $(1-p_{\rm about})\Delta\leq\varepsilon_{\rm sec}$, where $\|\cdot\|_{1}$ is the trace norm and $p_{\rm abort}$ is the probability that the protocol aborts. Therefore, we say that a protocol is $\varepsilon$-secure if it is $\varepsilon_{\textrm{cor}}$-correct and $\varepsilon_{\textrm{sec}}$-secret with $\varepsilon_{\textrm{cor}}+\varepsilon_{\textrm{sec}}\leq\varepsilon$.

\begin{table}[htb]
  \centering
  \begin{tabular}{ccccc}
   \hline
   \hline
  \quad & \multicolumn{4}{c}{Measurement results of Charlie}\\
  \hline
  & \multicolumn{2}{c}{Protocol 1} & \multicolumn{2}{c}{Protocol 2}\\
    \hline
  Alice \& Bob & $\ket{\Phi^-}$  & $\ket{\Psi^-}$ & $\ket{\Phi^-}$  &  $\ket{\Psi^-}$  \\
  \hline
  $\rm Z$ basis & No flip&   Flip   &  No flip  &  Flip \\
  $\rm X$ basis &   Flip &   Flip  & ---  & --- \\
  \hline\hline
\end{tabular}
\caption{Post-processing of raw key in the sifting step. Bob will decide whether he implements a key bit flip to guarantee correct correlations, depending on the announced entangled coherent state and the selected basis. Note that there is no key bit in the $\rm X$ basis for Protocol 2.
  }\label{table1}
\end{table}

\bigskip\noindent\textbf{Protocol definition.} Here, we follow two protocols proposed in our very recent work~\cite{Yin:2018:Twin}. One prepares cat state to bound the leaked information, called Protocol 1. The other exploits the phase-randomized coherent state (PRCS) to estimate the leaked information, called Protocol 2. For simplicity, we only consider the case of symmetric channel, while the case of the asymmetric channel can be directly generalized~\cite{Yin:2018:Twin}. The schematic diagram of two protocols are illustrated in Fig.~\ref{fig1}. Alice randomly chooses $\rm Z$ and $\rm X$ bases with probabilities $\ p_{\rm Z}$ and $1-p_{\rm Z}$, respectively. Alice randomly prepares optical pulses with coherent states $\ket{\alpha}$ and $\ket{-\alpha}$ in equal probabilities for the logic bits 0 and 1 if choosing the $\rm Z$ basis. For Protocol 1 (2), Alice randomly generates optical pulses with cat states $\ket{\xi^{+}(\alpha)}=(\ket{\alpha}+\ket{-\alpha})/\sqrt{2}$ and $\ket{\xi^{-}(\alpha)}=(\ket{\alpha}-\ket{-\alpha})/\sqrt{2}$ in equal probabilities for the logic bits 0 and 1 (PRCS) if choosing the $\rm X$ basis. Likewise, Bob does the same. The optical pulses are sent to the untrusted Charlie, who is assumed to perform the entangled coherent state measurement that projects them into an entangled coherent state. The decoy-state method~\cite{Hwang:2003:Quantum,wang2005beating,lo2005decoy} will be used in Protocol 2 to estimate the leaked information.

Next, Charlie will disclose whether he has acquired a successful measurement result and which entangled coherent state is obtained.
Alice and Bob only keep the data of successful measurement and discard the rest. They announce the basis and intensity information through the authenticated classical
channel and only keep the events of the same basis. Finally, Bob flips a part of his key bit to correctly correlate with Alice's (see Table~\ref{table1}). A detailed description of each step of Protocols 1 and 2 as follows.

\begin{description}
\item[1. State Preparation]
The first four steps are repeated by Alice and Bob for $i=1,\ldots,N$ until the conditions in the Sifting step are satisfied. 	
In Protocol 1, Alice chooses a basis $\beta\in\{\rm Z,\rm X\}$ and uniformly random bit $r\in\{0,1\}$ with probability $p_{\beta}/2$.
Next, Alice prepares optical pulses with coherent state $\ket{e^{ir\pi}\alpha}$ (cat state $(\ket{\alpha}+e^{ir\pi}\ket{-\alpha})/\sqrt{2}$) for $\rm Z$ ($\rm X$) basis given by $r$. Likewise, Bob does the same thing.
In Protocol 2, Alice chooses a basis $\beta\in\{\rm Z,\rm X\}$ with probability $p_{\beta}$. Then, she chooses uniformly random bit $r\in\{0,1\}$ with probability $1/2$ given by the $\rm Z$ basis and an intensity with probability $p_{a}$ given by the $\rm X$ basis.
Next, Alice prepares optical pulses with coherent state $\ket{e^{ir\pi}\alpha}$ for the $\rm Z$ basis given by $r$. She generates PRCS optical pulses of intensity $a$ for $\rm X$ basis.  Likewise, Bob does the same thing.

\item[2. Distribution] Alice and Bob send their optical pulses to untrusted Charlie through the insecure quantum channel.

\item[3. Measurement]
Charlie let the two optical pulses interfere in the symmetric beam splitter and performs the entangled state measurement.
For each $i$, he publicly informs Alice and Bob whether or not his measurement is successful and which entangled coherent state is obtained.

\item[4. Sifting]
Alice and Bob announce their basis choices and intensity settings over an authenticated classical channel when Charlie reports a successful event.
Bob flips part of his key bits to correctly correlate with Alice's (see Table \ref{table1}).
In Protocol 1, we define the set ${\mathcal Z}$ (${\mathcal X}$), which identifies
signals when Alice and Bob select the same basis $\rm Z$ ($\rm X$) and Charlie has a successful measurement. The protocol repeats these steps until $|{\mathcal Z}|\geq n$ and $|{\mathcal X}|\geq k$. In Protocol 2, we define two groups of sets ${\mathcal Z}$ and ${\mathcal X}_{a,b}$. The first (second) one identifies
signals where Alice and Bob select the basis $\rm Z$ ($\rm X$ and  the
intensities $a$ and $b$) and Charlie has a successful measurement.
The protocol repeats these steps till $|{\mathcal Z}|\geq n$ and $|{\mathcal X}_{a,b}|\geq k_{a,b}$ $\forall a, b$.

\item[5. Parameter Estimation]
Alice and Bob exploit the random bits from ${\mathcal Z}$ to form the raw key bit strings $\textbf{Z}$ and $\textbf{Z}'$, respectively. In Protocol 1 (2), Alice and Bob use ${\mathcal Z}$ and ${\mathcal X}$ (${\mathcal X_{a,b}}$) to estimate the upper bound of phase error rate $\phi_{\rm Z}$.  If $\phi_{\rm Z}>\phi_{\rm tol}$, Alice (Bob) assigns an empty string $\perp$ to $\textbf{S}$ ($\hat{\textbf{S}}$) and aborts this protocol.

\item[6. Error Correction] Bob exploits an information reconciliation scheme to acquire an estimate $\hat{\textbf{Z}}$
of $\textbf{Z}$ by revealing at most ${\rm leak}_{{\rm EC}}$ bits of error correction data.
Then, Alice computes a hash of length $\lceil\log_2(1/\epsilon_{\rm cor})\rceil$ by using a random universal$_2$ hash function~\cite{renner2008security} to $\textbf{Z}$. She sends the choice function and the hash to Bob. Bob uses the received hash function to compute the hash of $\hat{\textbf{Z}}$ and compares with Alice's. If they are different, Alice (Bob) assigns an empty string to $\textbf{S}$ ($\hat{\textbf{S}}$) and aborts this protocol.

\item[7. Privacy Amplification]  Alice exploits a random universal$_2$ hash function~\cite{renner2008security} to extract length $\ell$ bits of secret key $\textbf{S}$ from $\textbf{Z}$. Bob uses the same hash function (sent by Alice) to extract length $\ell$ bits of secret key $\hat{\textbf{S}}$ from $\hat{\textbf{Z}}$.
\end{description}

\begin{figure*}
\centering
\includegraphics[width=17cm]{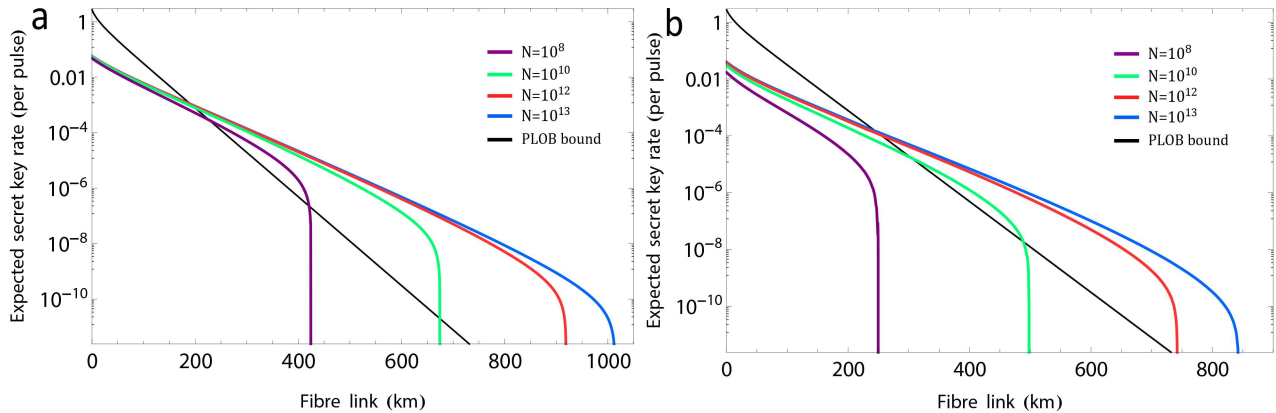}
\caption{Expected key rate as function of the distance. \textbf{a} (\textbf{b}), secret key rate $\ell/N$ in logarithmic scale for
Protocol 1 (2) as a function of the fibre distance. The colour lines correspond to different values for the total number of signals $N$ sent by Alice and Bob.
In comparison, the black line represents the repeaterless PLOB bound.
For simulation, we consider the following parameters: the loss coefficient of the fibre channel is $0.16$ dB/km, the detection efficiency and dark count rate are $85\%$ and $10^{-11}$. The overall misalignment rate in the channel is set to $2\%$, and the security bound of secrecy is $\varepsilon_{\rm sec}=10^{-10}$.
The results show clearly that the secret key rates of coherent-state-based TF-QKD in Protocols 1 and 2 can break the repeaterless PLOB bound even with a small finite size of data, say $N=10^{8}$ for Protocol 1 and $10^{10}$ for Protocol 2. The maximum transmission distance of Protocols 1 and 2 are more than 1000 km and 800 km with the realistic finite size of data $N=10^{13}$.
}\label{fig2}
\end{figure*}

Identifying any one of two entangled coherent states $\ket{\Phi^{-}}=1/\sqrt{N_{-}}(\ket{\alpha}\ket{\alpha}-\ket{-\alpha}\ket{-\alpha})$ and $\ket{\Psi^{-}}=1/\sqrt{N_{-}}(\ket{\alpha}\ket{-\alpha}-\ket{-\alpha}\ket{\alpha})$  can allow us to prove the security~\cite{Yin:2018:Twin}, where $N_{-}=2(1-e^{-4\mu})$ is the normalization factor, and $\mu=|\alpha|^{2}$ is the intensity of coherent states $\ket{\pm\alpha}$. Here, we consider that two entangled coherent states both can be identified. Indeed, the coherent-state-based TF-QKD is a prepare-and-measure protocol reduced from the entanglement-based QKD using heralded
entanglement generation protocol (see Methods).

\bigskip
\noindent
\textbf{Security analysis.}\label{sec_proof} Here, we show the main result of our paper. One can make sure that Protocol 1 (2) introduced above is both $\varepsilon_{\rm cor}$-correct and $\varepsilon_{\rm sec}$-secret if we choose an appropriate secret key of length $\ell$. The required correctness criterion could be ensured by the error-verification step. Alice and Bob compare the random hash values of their corrected keys with failure probability $\varepsilon_{\textrm{hash}}$, which means that identical probability of key bit strings $\textbf{S}$ and $\hat{\textbf{S}}$ is more than $1-\varepsilon_{\textrm{hash}}$. Even if the protocol is aborted, resulting in $\textbf{S}=\hat{\textbf{S}}=\perp$, it is also correct. Thereby, the correctness of the protocol is $\varepsilon_{\textrm{cor}}=\varepsilon_{\textrm{hash}}$.

For Protocol 1, the protocol is $\varepsilon_{\rm sec}$-secret if the secret key of length $\ell$ satisfies
\begin{equation}
\begin{aligned}\label{eq1}
\ell \leq n[1-h(\phi_{\rm Z})]-{\rm leak}_{{\rm EC}}-\log_{2}\frac{2}{\varepsilon_{\rm cor}}-2\log_{2}\frac{2}{\varepsilon_{\rm sec}},
\end{aligned}
\end{equation}
where $h(x)=-x\log_{2}x-(1-x)\log_{2}(1-x)$ is the binary Shannon entropy function. Recall that $n$ and $\phi_{\rm Z}$ are the number of bits and phase error rate in bit string $\textbf{Z}$. A sketch of the proof of Eq.~(\ref{eq1}) can be found in Methods.
In the asymptotic limit, $\phi_{\rm Z}=E_{\rm X}$ since statistical fluctuations could be neglected, and thus $\ell$ satisfies $\ell\leq n[1-h(E_{\rm X})]-{\rm leak}_{{\rm EC}}$, as recently acquired in \cite{Yin:2018:Twin}. $nh(\phi_{\rm Z})$ is the amount of information acquired by the eavesdropper in the quantum process, while ${\rm leak}_{\rm EC}$ is the information revealed by Alice in the error correction step.

For Protocol 2, the protocol is $\varepsilon_{\rm sec}$-secret if the secret key of length $\ell$ satisfies (see Methods)
\begin{equation}
\begin{aligned}\label{eq2}
\ell \leq n[1-h(\phi_{\rm Z})]-{\rm leak}_{{\rm EC}}-\log_{2}\frac{2}{\varepsilon_{\rm cor}}-2\log_{2}\frac{31}{2\varepsilon_{\rm sec}}.
\end{aligned}
\end{equation}

The other two main contributions of our work are the rigorous and tight statistical fluctuation analysis methods. One is the tightest multiplicative Chernoff bound and its variant to deal with the difference between the observed value and the expected value. The other is the tightest tail inequality for random sampling without replacement. In order to meet the composable security proof against general attacks in the finite-key regime, one can only assume the random variables are independent but not identically distributed. Traditionally, a large deviation theory with the Chernoff bound is proposed to deal with the parameter estimation in MDI-QKD with finite-key analysis~\cite{curty2014finite}, which is a rigorous but not tight method, i.e., significant statistical fluctuations quickly decrease the expected secret key rate in the high-loss regime.
Whereafter, another approach~\cite{zhang2017improved} is proposed, attempting to close the gap between the rigorous large-deviation Chernoff bound method~\cite{curty2014finite} and the not-sufficiently-rigorous Gaussian analysis (independent and identically distributed). However, this approach offers a tighter estimation of the lower bound (given the small observed value) than the Gaussian analysis, which seems to be a counterfactual result as the method~\cite{zhang2017improved} is superior to the Gaussian analysis. Our rigorously improved method are always inferior but comparable to the Gaussian analysis. Furthermore, we give two tailored tail inequalities (lower and upper tails) to deal with the random sampling without replacement issue, which directly utilizes hypergeometric function distribution and avoids any inequality scaling~\cite{Fung:2010:Practical,lim2014concise}. The rigorous proof and detailed analysis can be found in Supplementary Notes 1, 2 and 3.

\section*{Discussion}

\begin{figure*}
\centering
\includegraphics[width=17cm]{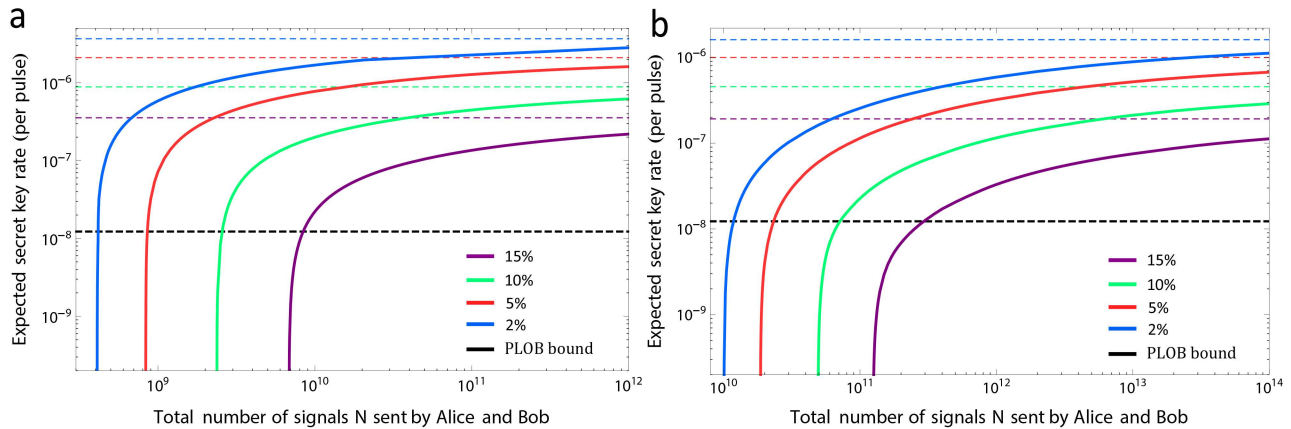}
\caption{Expected key rate as function of the block size. \textbf{a}, Protocol 1. \textbf{ b}, Protocol 2. The plot shows the secret key rate $\ell/N$ in logarithmic scale as a function of the total number of signals $N$
sent by Alice and Bob in the transmission distance of 500 km. The security bound of secrecy $\varepsilon_{\rm sec}=10^{-10}$.
The colour solid lines correspond to different values for the overall misalignment rate. The colour dotted lines show the corresponding asymptotic rates~\cite{Yin:2018:Twin}.
In comparison, the black line represents the PLOB bound given by the transmission distance of 500 km.
The results show that the coherent-state-based TF-QKD is robust to the large misalignment rate even for a finite size of signals sent by Alice and Bob.}\label{fig3}
\end{figure*}

Here, we perform the behaviour of the expected secret key rate provided in Eq.~(\ref{eq1}) of Protocol 1 and Eq.~(\ref{eq2}) of Protocol 2.
In our simulation, we use the following parameters, a fibre-based channel with an ultralow-loss of $0.16$ dB/km~\cite{Yin:2016:Measurement}.
The efficiency and dark count rate of single-photon detector are $85\%$ and $10^{-11}$ in the untrusted relay~\cite{Boaron:2018:Secure}.
The security bounds of secrecy and correctness are fixed to $\varepsilon_{\rm sec}=10^{-10}$ and $\varepsilon_{\rm cor}=10^{-15}$, the latter of which corresponds to a realistic
hash tag size in practice~\cite{renner2008security}. For simplicity, we assume an error correction leakage that is a fixed fraction of the sifted key length $n$, i.e.,
${\rm leak}_{\rm EC}=n\zeta{}h(E_{\rm Z})$, with the efficiency of error correction $\zeta=1.1$ and the quantum bit error rate $E_{\rm Z}$ of the ${\rm Z}$ basis.

The results are shown in Figs. \ref{fig2} and \ref{fig3} where Alice and Bob exploit the three-intensity PRCS, one of which is a vacuum state. The detailed computational process of the phase error rate $\phi_{\rm Z}$ can be found in Methods.
The expected secret key rate (per pulse) $\ell/N$ as a function of the transmission distance between Alice and Bob for different values of the total number of signals $N$ sent by Alice and Bob given by overall misalignment $2\%$ in the channel is shown in Fig 2. For a given transmission distance, we optimize numerically $\ell/N$ over all the free parameters of Protocols 1 and 2. For the case of symmetric channel, all parameters chosen by Alice and Bob are set to the same.
Our simulation result shows clearly that coherent-state-based TF-QKD is the feasible scheme in the finite-key regime.
Considering the case of 1 GHz repetition rate~\cite{Minder:2019:Experimental}, the secret key rate of Protocols 1 and 2 can break the repeaterless Pirandola-Laurenza-Ottaviani-Banchi (PLOB) bound~\cite{pirandola2017fundamental} even with a small finite size of data, say $N=10^{8}$ (data collected in 0.1 s) for Protocol 1 and $10^{10}$ (data collected in 10 s) for Protocol 2. Moreover, the maximum transmission distance of Protocols 1 and 2 can be expanded up to 1000 km and 800 km with the realistic finite size of data $N=10^{13}$ (less than 2.8 h data). The secret key rate in Protocols 1 and 2 given by 470 km are both larger than $10^{-6}$ per pulse (1 kbps) under the finite size of data $N=10^{12}$. It means that the coherent-state-based TF-QKD has the potential to be actually used even when the communication distance is approximate to 500 km. This is impossible when using the traditional QKD or MDI-QKD, where the best results are 0.25 bps at 421 km of traditional QKD under the collective attacks assumption~\cite{Boaron:2018:Secure} and $3.2\times10^{-4}$ bps at 404 km of MDI-QKD under the coherent attacks assumption~\cite{Yin:2016:Measurement}.

Figure 3 illustrates $\ell/N$ as a function of $N$ for different values of the misalignment in the transmission distance of 500 km.
For comparison, this figure also includes the asymptotic secret key rate when Alice and Bob send an infinite number of signals \cite{Yin:2018:Twin} and the repeaterless PLOB bound.  For a given number of signals, we optimize numerically $\ell/N$ over all the free parameters of Protocols 1 and 2. The fixed parameters are the ones described in the caption of Fig.~\ref{fig3}. The simulation results show that the secret key rates of Protocols 1 and 2 are about $10^{-7}$ at the distance of 500 km with $10^{11}$ and $5\times10^{13}$ signals, even given that the misalignment rate is up to $15\%$. The significant secret key rate of Protocols 1 and 2 at the distance of 500 km can be acquired only with $10^{9}$ and $10^{11}$ signals when the misalignment rate is less than $5\%$.

In summary, we have proved the composable security of coherent-state-based TF-QKD in the finite-key regime against general attacks. The maximum transmission distance of Protocols 1 and 2 are more than 1000 km and 800 km with the realistic finite size of data, respectively.
The coherent-state-based TF-QKD is the fully practical QKD protocol that offers an avenue to bridge the gap between trusted relay and quantum repeater in long-distance QKD implementations. In order to be immune to general attacks in the finite-key regime, the independent and identically distributed assumption of Gaussian analysis (the central-limit theorem) is no longer applicable.
We have rigorously proved an improved Chernoff bound and its variant, which can close the gap between the large-deviation Chernoff bound method and the Gaussian analysis. Numerical simulations display that our improved method is always inferior but comparable to the Gaussian analysis. The rigorous and tight statistical fluctuation analysis methods of this work will be widely applied to quantum cryptography protocols with the finite-size effects, such as QKD, quantum digital signature, and quantum secret sharing. We remark that cat state has a certain distance from the actual application with current technique.
Last but not least, the homodyne measurement may be exploited to identify the entangled coherent state in the coherent-state-based TF-QKD, which is worth considering in the future.

\section*{Methods}
\noindent
\textbf{Entanglement-based protocol.}\label{ent}
In order to establish the secrecy of the protocols, we introduce an equivalently virtual entanglement-based protocol~\cite{Yin:2018:Twin}, in which
Alice and Bob prepare entangled states of a qubit and an optical mode $\ket{\psi}=\frac{1}{\sqrt{2}}\left(\ket{+z}\ket{\alpha}+\ket{-z}\ket{-\alpha}\right)$,
where qubit states $\ket{\pm z}$ are the eigenstates of Pauli's $Z$ operator. They keep the qubit and send the optical mode to the untrusted Charlie, who performs the entangled coherent state measurement. The bipartite qubit entanglement states between Alice and Bob are thus generated via entanglement swapping. Indeed,  the coherent states $\ket{\pm\alpha}$ and the cat states $\ket{\xi^{\pm}(\alpha)}$ will be sent to Charlie if they perform the $Z$- and $X$-basis measurement on the qubit system, respectively. Thereby, the coherent-state-based TF-QKD is a prepare-and-measure protocol reduced from the entanglement-based QKD using heralded entanglement generation protocol  (we refer to the article~\cite{Yin:2018:Twin} for details).

\bigskip
\noindent
\textbf{Secrecy.}\label{sec} Let us keep the entanglement-based QKD using heralded entanglement generation protocol in our mind. We exploit the entropic uncertainty relations~\cite{tomamichel2011uncertainty,tomamichel2012tight} to estimate bounds on the smooth min-entropy of the raw key conditioned on eavesdropper's information.
The Quantum Leftover Hash Lemma~\cite{renner2008security} is exploited to give a direct operational meaning to the smooth min-entropy. Let $\textbf{E}'$ summarizes all information of eavesdropper learned about raw key of Alice $\textbf{Z}$, up to the error-correction step. By applying a random universal$_{2}$ hash function to $\textbf{Z}$, one may extract a $\Delta$-secret key of length $\ell$ from $\textbf{Z}$,
\begin{equation}
\begin{aligned}\label{eq}
\Delta=2\epsilon+\frac{1}{2}\sqrt{2^{\ell-H_{\rm min}^{\epsilon}(\textbf{Z}|\textbf{E}')}},
\end{aligned}
\end{equation}
where $H_{\rm min}^{\epsilon}(\textbf{Z}|\textbf{E}')$ denotes the smooth min-entropy~\cite{renner2008security}, which quantifies the average probability that the eavesdropper guesses $\textbf{Z}$ correctly by exploiting the optimal strategy with access to $\textbf{E}'$. Let $\upsilon=\sqrt{2^{\ell-H_{\rm min}^{\epsilon}(\textbf{Z}|\textbf{E}')}}/2$, the secret key of length $\ell$ is
\begin{equation}
\begin{aligned}\label{eq}
\ell=\bigg\lfloor H_{\rm min}^{\epsilon}(\textbf{Z}|\textbf{E}')-2\log_{2}\frac{1}{2\upsilon}\bigg\rfloor.
\end{aligned}
\end{equation}

The amount of bit information ${\rm leak}_{{\rm EC}}+\log_{2}(2/\varepsilon_{\rm cor})$ will be revealed to the adversary during the error-correction step. By using a chain-rule inequality for smooth entropies, we have
$H_{\rm min}^{\epsilon}(\textbf{Z}|\textbf{E}')\geq H_{\rm min}^{\epsilon}(\textbf{Z}|\textbf{E})-{\rm leak}_{{\rm EC}}-\log_{2}(2/\varepsilon_{\rm cor})$, where $\textbf{E}$ is the information of eavesdropper before the classical post-processing.

In order to bound the smooth min-entropy $H_{\rm min}^{\epsilon}(\textbf{Z}|\textbf{E})$ by using the uncertainty relation for smooth entropies~\cite{tomamichel2011uncertainty}, we consider a gedankenexperiment that Alice and Bob prepare the cat states instead of coherent states when they choose the $Z$ basis. Alice and Bob need to use the bit strings $\textbf{X}$ and $\textbf{X}'$ of length $n$ to replace the raw key bit strings $\textbf{Z}$ and $\textbf{Z}'$ in this hypothetical protocol, respectively. The smooth min-entropy can be given by
\begin{equation}
\begin{aligned}\label{eq}
H_{\rm min}^{\epsilon}(\textbf{Z}|\textbf{E})&\geq n-H_{\rm max}^{\epsilon}(\textbf{X}|\textbf{X}')\\
&=n[1-h(\phi_{\rm Z})],
\end{aligned}
\end{equation}
where the first inequality exploits the entropic uncertainty relation~\cite{tomamichel2011uncertainty}. The smooth max-entropy
$H_{\rm max}^{\epsilon}(\textbf{X}|\textbf{X}')$ quantifies the required number of bits that Bob uses bit string $\textbf{X}'$ to reconstruct $\textbf{X}$, which leads to the second inequality~\cite{tomamichel2012tight}. $\phi_{\rm Z}$ is the phase error rate of bit strings $\textbf{Z}$ and $\textbf{Z}'$, i.e., the bit error rate of bit strings $\textbf{X}$ and $\textbf{X}'$. In reality, $\phi_{Z}$ cannot be directly observed, which has to be estimated by using random-sampling (without replacement) theory.

\bigskip
\noindent
\textbf{Tight tail inequality.}\label{ent} Here, we introduce three Lemmas to deal with the statistical fluctuation in the finite-key regime. Specifically, Lemma 1 is  tailored for random sampling without replacement. Lemma 2 is the multiplicative Chernoff bound, which is used to bound the observed value, given the expected value. Lemma 3 is a variant of the multiplicative Chernoff bound, which is tailored to estimate the expected value, given the observed value. The rigorously proved tail inequalities in each lemma are the tightest due to avoiding excessive inequality scaling. See Supplementary Notes 1, 2 and 3 for details.

Lemma 1: Let ${\mathcal X_{n+k}}:=\{x_{1},x_{2},\cdots,x_{n+k}\}$ be a string of binary bits with $n+k$ size, in which the number of bit value 1 is unknown. Let ${\mathcal X_{k}}$ be a random sample (without replacement) bit string with $k$ size from ${\mathcal X_{n+k}}$. Let $\lambda_{k}$ be the probability of observed bit value 1 in ${\mathcal X_{k}}$. Let ${\mathcal X_{n}}$ be the remaining bit string, where the probability of observed bit value 1 in ${\mathcal X_{n}}$ is $\lambda_{n}$.
Then, let $C_{i}^{j}=i!/[j!(i-j)!]$ be the  binomial coefficient. For any $\epsilon>0$, we have the upper tail
\begin{equation}
\begin{aligned}\label{eq}
{\rm Pr}[\lambda_{n}\geq \lambda_{k}+\gamma(n,k,\lambda_{k},\epsilon)]\leq\epsilon,
\end{aligned}
\end{equation}
where $\gamma(a,b,c,d)$ is the positive root of the equation
$\ln C_{b}^{bc}+\ln C_{a}^{ac+a\gamma(a,b,c,d)}-\ln C_{a+b}^{(a+b)c+a\gamma(a,b,c,d)}-\ln d=0$.
For any $\hat{\epsilon}>0$, we have the lower tail
\begin{equation}
\begin{aligned}\label{eq}
{\rm Pr}[\lambda_{n}\leq \lambda_{k}-\hat{\gamma}(n,k,\lambda_{k},\hat{\epsilon})]\leq\hat{\epsilon},
\end{aligned}
\end{equation}
where $\hat{\gamma}(a,b,c,d)$ is the positive root of the equation
$\ln C_{b}^{bc}+\ln C_{a}^{ac-a\hat{\gamma}(a,b,c,d)}-\ln C_{a+b}^{(a+b)c-a\hat{\gamma}(a,b,c,d)}-\ln d=0$.
If one does not find the positive root $\hat{\gamma}(a,b,c,d)$, we let $\lambda_{n}=0$.

Lemma 2: Let $X_1, X_2...,X_N$ be a set of independent Bernoulli random variables that satisfy ${\rm Pr}(X_i=1)=p_i$ (not necessarily equal), and let $X:=\sum_{i=1}^NX_i$.
The expected value of $X$ is denoted as $\mu_{x}:=E[X]=\sum_{i=1}^Np_i$. Then, let $g(x,y)=\left[e^{y}/(1+y)^{1+y}\right]^{x}$, for any $\delta>0$, we have the upper tail
\begin{equation}
\begin{aligned}\label{eq}
{\rm Pr}[X\geq(1+\delta)\mu_{x}]<g(\mu_{x},\delta)=\epsilon,
\end{aligned}
\end{equation}
where $\delta$ is the positive root of the equation $\mu_{x}[\delta-(1+\delta)\ln(1+\delta)]-\ln\epsilon=0$.
For any $0<\hat{\delta}\leq1$, we have the lower tail
\begin{equation}
\begin{aligned}\label{eq}
{\rm Pr}[X\leq(1-\hat{\delta})\mu_{x}]<g(\mu_{x},-\hat{\delta})=\hat{\epsilon},
\end{aligned}
\end{equation}
where $\hat{\delta}$ is the positive root of the equation $\mu_{x}[\hat{\delta}+(1-\hat{\delta})\ln(1-\hat{\delta})]+\ln\hat{\epsilon}=0$.

Lemma 3: Let $X_1, X_2...,X_N$ be a set of independent Bernoulli random variables that satisfy ${\rm Pr}(X_i=1)=p_i$ (not necessarily equal), and let $X:=\sum_{i=1}^NX_i$.
The expected value of $X$ is denoted as $\mu_{x}:=E[X]=\sum_{i=1}^Np_i$. An observed outcome of $X$ is represented as $x$ for a given trial.
For any $\epsilon>0$, we have $\mu_{x}$ that satisfies
\begin{equation}
\begin{aligned}\label{eq}
\mu_{x}\geq\underline{\mu_{x}}= \max\{0,x-\Delta(x,\epsilon)\},
\end{aligned}
\end{equation}
with failure probability $\epsilon$, where $\underline{\mu_{x}}$ is the lower bound of $\mu_{x}$ and $\Delta(z,y)$ is the positive root of the equation
$\Delta(z,y)-[z+\Delta(z,y)]\ln[1+\Delta(z,y)/z]-\ln y=0$.
For any $\hat{\epsilon}>0$, we have that $\mu_{x}$ satisfies
\begin{equation}
\begin{aligned}\label{eq}
\mu_{x}\leq\overline{\mu_{x}}= x+\hat{\Delta}(x,\hat{\epsilon}),
\end{aligned}
\end{equation}
with failure probability $\hat{\epsilon}$, where $\overline{\mu_{x}}$ is the upper bound of $\mu_{x}$ and $\hat{\Delta}(z,y)$ is the positive root of the equation
$\hat{\Delta}(z,y)+z\ln\{z/[z+\hat{\Delta}(z,y)]\}+\ln y=0$.

\bigskip
\noindent
\textbf{Statistical fluctuation of Protocol 1.}\label{ent}
In order to bound the phase error rate $\phi_{\rm Z}$, we consider the gedankenexperiment picture. There are $n+k$ bits corresponding to $\rm X$ basis. The observed error rate of $k$ bits random sampled from $n+k$ bits is $E_{\rm X}=\frac{1}{k}\sum_{j=1}^{k} r_{x}\oplus r'_{x}$,
where $r_{x}$ and $r'_{x}$ are Alice's and Bob's bits in set ${\mathcal X}$. By using the upper tail inequality for random sampling without replacement in Lemma 1, the remaining error rate of $n$ bits, i.e., the phase error rate, can be given by
\begin{equation}
\begin{aligned}\label{eq}
\phi_{\rm Z}\leq E_{\rm X}+\gamma(n,k,E_{\rm X},\epsilon_{1}),
\end{aligned}
\end{equation}
with failure probability $\epsilon_{1}$.

Finally, by composing the failure probability due to parameter estimation, we have a total secrecy of $\varepsilon_{\rm sec}=2\epsilon+\upsilon+\epsilon_{1}$, where we take $\epsilon=\upsilon=\epsilon_{1}=\varepsilon_{\rm sec}/4$.

\bigskip
\noindent
\textbf{Statistical fluctuation of Protocol 2.}\label{ent}
Since the cat states are replaced by PRCS for the ${\rm X}$ basis choice in Protocol 2, the bit error rate $E_{\rm X}$ in the ${\rm X}$ basis cannot be directly observed.
In order to bound the phase error rate $\phi_{\rm Z}$, we need to use the following three steps.

First, let $Q_{a,b}^{*}$ be the expected gain when Alice and Bob send PRCS with intensities $a$ and $b$, respectively, $a,b\in\{\nu,\omega,0\}$. Therefore, we have the relations $k_{a,b}^{*}=Np_{\rm X}^{2} p_{a}p_{b}Q_{a,b}^{*}$, where $k_{a,b}^{*}$ are the expected values corresponding to the observed values $k_{a,b}$. In reality, we only know the observed values $k_{a,b}$. By using a variant of the multiplicative Chernoff bound in Lemma 3, we can use the observed value for a given trial to estimate the upper (lower) bound of the expected value with a small failure probability $\epsilon_{3}$. The PRCS can be seen as the mixed Fock states from the eavesdropper's view. Let $Y_{n,m}^{*}$ be the expected yield when Alice sends $n$-photon and Bob sends $m$-photon.
Thereby, the expected values $Y_{n,m}^{*}$ can be estimated by using the decoy-state method with the three-intensity PRCS~\cite{cui:2018:phase,grasselli2019practical,Yin:2018:Twin}. Once obtaining the upper bound of the expected yield $\overline{Y}_{n,m}^{*}$, one can calculate the upper bound of the observed yield $\overline{Y}_{n,m}$  by using the lower tail of the multiplicative Chernoff bound in Lemma 2. See Supplementary Note 4 for details. Note that for the case of $n+m\geq5$, we let the observed yield $\overline{Y}_{n,m}=1$.

Second, we consider the gedankenexperiment picture, in which Alice and Bob still send the cat states $\ket{\xi^{\pm}(\alpha)}$ instead of PRCS when they choose the ${\rm X}$ basis in Protocol 2. Let $Q_{\rm Z}$ ($Q_{\rm X}$) be the observed gain when Alice and Bob both prepare coherent states $\ket{\pm\alpha}$ (cat states $\ket{\xi^{\pm}(\alpha)}$) for a given trial. By using the tail inequality for random sampling without replacement in Lemma 1, the observed value $Q_{\rm X}$ can be bounded by
\begin{equation}
\begin{aligned}\label{}
Q_{\rm X}\geq \underline{Q}_{\rm X}=Q_{\rm Z}-\hat{\gamma}(N_{\rm X},N_{\rm Z},Q_{\rm Z},\epsilon_{1}),
\end{aligned}
\end{equation}
with failure probability $\epsilon_{1}$, where we have the relations $n=N_{\rm Z}Q_{\rm Z}$, $N_{\rm Z}=Np_{\rm Z}^{2}$ and $N_{\rm X}=Np_{\rm X}^{2}$.
Thereby, the lower bound of the observed value is $\underline{k}=N_{\rm X}\underline{Q}_{\rm X}$.

Third, the upper bound of the observed value of the bit error rate $\overline{E}_{\rm X}$ can be estimated by~\cite{Yin:2018:Twin}
\begin{equation}
\begin{aligned}\label{eq}
\overline{E}_{\rm X}\leq \overline{Q}_{\rm X}^{E}/\underline{Q}_{\rm X},
\end{aligned}
\end{equation}
where we have the error gain~\cite{curty:2018:simple,cui:2018:phase,Yin:2018:Twin}
\begin{equation}
\begin{aligned}\label{eq}
\overline{Q}_{\rm X}^{E}&\leq \left(\sum_{n,m=0}^{\infty}\sqrt{P_{2n}^{\mu}P_{2m}^{\mu}\overline{Y}_{2n,2m}}\right)^{2}\\
&+\left(\sum_{n,m=0}^{\infty}\sqrt{P_{2n+1}^{\mu}P_{2m+1}^{\mu}\overline{Y}_{2n+1,2m+1}}\right)^{2},
\end{aligned}
\end{equation}
with $P_{n}^{\mu}=e^{-\mu}\mu^{n}/n!$.
By using the upper tail inequality for random sampling without replacement in Lemma 1, the phase error rate can be given by
\begin{equation}
\begin{aligned}\label{eq}
\phi_{\rm Z}\leq \overline{E}_{\rm X}+\gamma(n,\underline{k},\overline{E}_{\rm X},\epsilon_{1})\},
\end{aligned}
\end{equation}
with failure probability $\epsilon_{1}$. We remark that the joint constraint method~\cite{zhou2016making} will further bound phase error rate in the finite-key regime.

Finally, by composing the failure probability due to parameter estimation, we have a total secrecy of $\varepsilon_{\rm sec}=2\epsilon+\upsilon+2\epsilon_{1}+9\epsilon_{2}+17\epsilon_{3}$, where we take $\epsilon=\upsilon=\epsilon_{1}=\epsilon_{2}=\epsilon_{3}=\varepsilon_{\rm sec}/31$.

\noindent\textbf{Acknowledgments}\\
We thank Y. Fu, P. Liu and W. Zhu for their valuable discussions. This work was supported by the National Natural Science Foundation of China under Grant No. 61801420 and the Nanjing University.

\noindent\textbf{Author Contributions}\\
H.-L.Y. and Z.-B.C. conceived and designed the study. H.-L.Y. performed the numerical simulation.
All authors contributed extensively to the work presented in this paper.

\noindent\textbf{Additional Information}\\
Competing interests: The authors declare no competing interests.

\section*{References}


\bibliographystyle{naturemag}

%

\newpage
\onecolumngrid

\section*{Supplementary Note 1: Random sampling without replacement.}
Here, we present the proof for the lemma of random sampling without replacement in the Methods of the main text.
The new tail inequality of random sampling without replacement is the tightest due to avoiding any inequality scaling.

\noindent
\textbf{Lemma 1.} Tight tail inequality of random sampling without replacement.

Let ${\mathcal X_{n+k}}:=\{x_{1},x_{2},\cdots,x_{n+k}\}$ be a string of binary bits with $n+k$ size, in which the number of bit value 1 is unknown. Let ${\mathcal X_{k}}$ be a random sample (without replacement) bit string with $k$ size from ${\mathcal X_{n+k}}$. Let $\lambda_{k}$ be the probability of bit value 1 observed in ${\mathcal X_{k}}$. Let ${\mathcal X_{n}}$ be the remaining bit string, where the probability of bit value 1 observed in ${\mathcal X_{n}}$ is $\lambda_{n}$.
Then let $C_{i}^{j}=i!/[j!(i-j)!]$ be the  binomial coefficient. For any $\epsilon>0$, we have the upper tail
\begin{equation}
\begin{aligned}\label{eq1}
{\rm Pr}[\lambda_{n}\geq \lambda_{k}+\gamma(n,k,\lambda_{k},\epsilon)]\leq\epsilon,
\end{aligned}
\end{equation}
where $\gamma(a,b,c,d)$ is the positive root of the following equation
\begin{equation}
\begin{aligned}\label{eq2}
\ln C_{b}^{bc}+\ln C_{a}^{ac+a\gamma(a,b,c,d)}-\ln C_{a+b}^{(a+b)c+a\gamma(a,b,c,d)}-\ln d=0.
\end{aligned}
\end{equation}
For any $\hat{\epsilon}>0$, we have the lower tail
\begin{equation}
\begin{aligned}\label{eq3}
{\rm Pr}[\lambda_{n}\leq \lambda_{k}-\hat{\gamma}(n,k,\lambda_{k},\hat{\epsilon})]\leq\hat{\epsilon},
\end{aligned}
\end{equation}
where $\hat{\gamma}(a,b,c,d)$ is the positive root of the following equation
\begin{equation}
\begin{aligned}\label{eq4}
\ln C_{b}^{bc}+\ln C_{a}^{ac-a\hat{\gamma}(a,b,c,d)}-\ln C_{a+b}^{(a+b)c-a\hat{\gamma}(a,b,c,d)}-\ln d=0.
\end{aligned}
\end{equation}
If one does not find the positive root $\hat{\gamma}(a,b,c,d)$, we let $\lambda_{n}=0$.

\noindent
\textbf{Proof.}

First, we prove the inequality of the upper tail. Let $X=n \lambda_{n}+k \lambda_{k}$, we have

\begin{equation}
\begin{aligned}\label{eq5}
{\rm Pr}[\lambda_{n}\geq \lambda_{k}+\gamma]&={\rm Pr}[X\geq (n+k)\lambda_{k}+n\gamma,k\lambda_{k}]\\
&=\sum_{X=(n+k)\lambda_{k}+n\gamma}^{n+k\lambda_{k}}{\rm Pr}[X,k\lambda_{k}]\\
&=\sum_{X=(n+k)\lambda_{k}+n\gamma}^{n+k\lambda_{k}}{\rm Pr}[k\lambda_{k}|X]{\rm Pr}[X]\\
&=\sum_{X=(n+k)\lambda_{k}+n\gamma}^{n+k\lambda_{k}}\frac{C_{k}^{k \lambda_{k}}C_{n}^{X-k\lambda_{k}}}{C_{n+k}^{X}}{\rm Pr}[X]\\
&\leq\frac{C_{k}^{k \lambda_{k}}C_{n}^{n\lambda_{k}+n\gamma}}{C_{n+k}^{(n+k)\lambda_{k}+n\gamma}},
\end{aligned}
\end{equation}
where we use the fact that the conditional probability ${\rm Pr}[k\lambda_{k}|X]=C_{k}^{k \lambda_{k}}C_{n}^{X-k\lambda_{k}}/C_{n+k}^{X}$ is the hypergeometric distribution function and is a monotonic decreasing function of $X$ when $X\geq(n+k)\lambda_{k}$. By using Eq. \eqref{eq2}, we find
\begin{equation}
\begin{aligned}\label{eq6}
\frac{C_{k}^{k \lambda_{k}}C_{n}^{n\lambda_{k}+n\gamma(n,k,\lambda_{k},\epsilon)}}{C_{n+k}^{(n+k)\lambda_{k}+n\gamma(n,k,\lambda_{k},\epsilon)}}=\epsilon.
\end{aligned}
\end{equation}
Thereby, we have proved the upper tail ${\rm Pr}[\lambda_{n}\geq \lambda_{k}+\gamma(n,k,\lambda_{k},\epsilon)]\leq\epsilon$.

Now, we prove the inequality of the lower tail. We consider the case of $\lambda_{k}\geq\hat{\gamma}(n,k,\lambda_{k},\hat{\epsilon})\geq0$. Let $\hat{X}=n \lambda_{n}+k \lambda_{k}$, we have

\begin{equation}
\begin{aligned}\label{eq7}
{\rm Pr}[\lambda_{n}\leq \lambda_{k}-\hat{\gamma}]&={\rm Pr}[\hat{X}\leq (n+k)\lambda_{k}-n\hat{\gamma},k\lambda_{k}]\\
&=\sum_{\hat{X}=k\lambda_{k}}^{(n+k)\lambda_{k}-n\hat{\gamma}}{\rm Pr}[\hat{X},k\lambda_{k}]\\
&=\sum_{\hat{X}=k\lambda_{k}}^{(n+k)\lambda_{k}-n\hat{\gamma}}{\rm Pr}[k\lambda_{k}|\hat{X}]{\rm Pr}[\hat{X}]\\
&=\sum_{\hat{X}=k\lambda_{k}}^{(n+k)\lambda_{k}-n\hat{\gamma}}\frac{C_{k}^{k \lambda_{k}}C_{n}^{\hat{X}-k\lambda_{k}}}{C_{n+k}^{\hat{X}}}{\rm Pr}[\hat{X}]\\
&\leq\frac{C_{k}^{k \lambda_{k}}C_{n}^{n\lambda_{k}-n\hat{\gamma}}}{C_{n+k}^{(n+k)\lambda_{k}-n\hat{\gamma}}},
\end{aligned}
\end{equation}
where we use the fact that the conditional probability ${\rm Pr}[k\lambda_{k}|\hat{X}]=C_{k}^{k \lambda_{k}}C_{n}^{\hat{X}-k\lambda_{k}}/C_{n+k}^{\hat{X}}$ is the hypergeometric distribution function and is a monotonic increasing function of $\hat{X}$ when $\hat{X}\leq(n+k)\lambda_{k}$. By using Eq. \eqref{eq4}, we find
\begin{equation}
\begin{aligned}\label{eq8}
\frac{C_{k}^{k \lambda_{k}}C_{n}^{n\lambda_{k}-n\hat{\gamma}(n,k,\lambda_{k},\hat{\epsilon})}}{C_{n+k}^{(n+k)\lambda_{k}-n\hat{\gamma}(n,k,\lambda_{k},\hat{\epsilon})}}=\hat{\epsilon}.
\end{aligned}
\end{equation}
Thereby, we have proved the lower tail ${\rm Pr}[\lambda_{n}\leq \lambda_{k}-\hat{\gamma}(n,k,\lambda_{k},\hat{\epsilon})]\leq\hat{\epsilon}$.

\section*{Supplementary Note 2: The multiplicative Chernoff bound and its variant.}
Here, we give the proof for the Lemma of the multiplicative Chernoff bound and its variant shown in the Methods of the main text.
First, we prove that the multiplicative Chernoff bound is almost the tightest.  The multiplicative Chernoff bound is exploited to estimate the observed value, given the expected value. Second, we propose a variant of the multiplicative Chernoff bound as tight as possible which is used to bound the expected value, given the observed value.

\noindent
\textbf{Lemma 2.} Tight multiplicative Chernoff bound.

Let $X_1, X_2...,X_N$ be a set of independent Bernoulli random variables that satisfy ${\rm Pr}(X_i=1)=p_i$ (not necessarily equal), and let $X:=\sum_{i=1}^NX_i$.
The expected value of $X$ is denoted as $\mu_{x}:=E[X]=\sum_{i=1}^Np_i$. Then, let $g(x,y)=\left[\frac{e^{y}}{(1+y)^{1+y}}\right]^{x}$, for any $\delta>0$, we have the upper tail
\begin{equation}
\begin{aligned}\label{eq}
{\rm Pr}[X\geq(1+\delta)\mu_{x}]<g(\mu_{x},\delta)=\epsilon,
\end{aligned}
\end{equation}
where $\delta$ is the positive root of the following equation
\begin{equation}
\begin{aligned}\label{equ}
\mu_{x}[\delta-(1+\delta)\ln(1+\delta)]-\ln\epsilon=0.
\end{aligned}
\end{equation}
For any $0<\hat{\delta}\leq1$, we have the lower tail
\begin{equation}
\begin{aligned}\label{eq}
{\rm Pr}[X\leq(1-\hat{\delta})\mu_{x}]<g(\mu_{x},-\hat{\delta})=\hat{\epsilon},
\end{aligned}
\end{equation}
where $\hat{\delta}$ is the positive root of the following equation
\begin{equation}
\begin{aligned}\label{eql}
\mu_{x}[\hat{\delta}+(1-\hat{\delta})\ln(1-\hat{\delta})]+\ln\hat{\epsilon}=0.
\end{aligned}
\end{equation}

\noindent
\textbf{Proof.}

First, we prove the first inequality of upper tail. For $t>0$, we can have an equivalent inequality,
\begin{equation}
\begin{aligned}\label{eq11}
{\rm Pr}[X\geq(1+\delta)\mu_{x}]={\rm Pr}[e^{tX}\geq e^{t(1+\delta)\mu_{x}}].
\end{aligned}
\end{equation}
By exploiting the Markov inequality, the above inequality can be given by
\begin{equation}
\begin{aligned}\label{eq12}
{\rm Pr}[X\geq(1+\delta)\mu_{x}]={\rm Pr}[e^{tX}\geq e^{t(1+\delta)\mu_{x}}]\leq\frac{E[e^{tX}]}{e^{t(1+\delta)\mu_{x}}}.
\end{aligned}
\end{equation}
Since $X=\Sigma_{i=1}^{N}X_{i}$, we have $E[e^{tX}]=\Pi_{i=1}^{N}E[e^{tX_{i}}]$. The independent Bernoulli random variables satisfy ${\rm Pr}(X_i=1)=p_i$. The expected value is $E[e^{tX_{i}}]=1+p_{i}(e^{t}-1)<e^{p_{i}(e^{t}-1)}$, where we use the fact that $e^{y}>(1+y)$ for $y>0$.
Thereby, we have the inequality
\begin{equation}
\begin{aligned}\label{eq13}
E[e^{tX}]=\prod_{i=1}^{N}E[e^{tX_{i}}]<\prod_{i=1}^{N}e^{p_{i}(e^{t}-1)}=e^{\sum_{i=1}^{N}p_{i}(e^{t}-1)}=e^{(e^{t}-1)\mu_{x}}.
\end{aligned}
\end{equation}
Substituting Eq. \eqref{eq13} back into Eq. \eqref{eq12}, the final inequality can be bounded by
\begin{equation}
\begin{aligned}\label{eqll4}
{\rm Pr}[X\geq(1+\delta)\mu_{x}]<\frac{e^{(e^{t}-1)\mu_{x}}}{e^{t(1+\delta)\mu_{x}}}=\left[\frac{e^{\delta}}{(1+\delta)^{1+\delta}}\right]^{\mu_{x}},
\end{aligned}
\end{equation}
where we assume $t=\ln(1+\delta)$ to \textbf{make the bound as tight as possible}. By using Eq.\eqref{equ}, we have
\begin{equation}
\begin{aligned}\label{eqll4}
{\rm Pr}[X\geq(1+\delta)\mu_{x}]<=\left[\frac{e^{\delta}}{(1+\delta)^{1+\delta}}\right]^{\mu_{x}}=g(\mu_{x},\delta)=\epsilon.
\end{aligned}
\end{equation}

Now, we prove the second inequality of lower tail by using the similar method. For $t>0$, we have an equivalent inequality as follows,
\begin{equation}
\begin{aligned}\label{eq15}
{\rm Pr}[X\leq(1-\hat{\delta})\mu_{x}]={\rm Pr}[e^{-tX}\geq e^{-t(1-\hat{\delta})\mu_{x}}].
\end{aligned}
\end{equation}
The above inequality can be bounded by the Markov inequality,
\begin{equation}
\begin{aligned}\label{eq16}
{\rm Pr}[X\leq(1-\hat{\delta})\mu_{x}]={\rm Pr}[e^{-tX}\geq e^{-t(1-\hat{\delta})\mu_{x}}]\leq\frac{E[e^{-tX}]}{e^{-t(1-\hat{\delta})\mu_{x}}}.
\end{aligned}
\end{equation}
Obviously, $E[e^{-tX}]=\Pi_{i=1}^{N}E[e^{-tX_{i}}]$ because $X=\Sigma_{i=1}^{N}X_{i}$. The expected value is $E[e^{-tX_{i}}]=1+p_{i}(e^{-t}-1)<e^{p_{i}(e^{-t}-1)}$ since the independent Bernoulli random variables satisfy ${\rm Pr}(X_i=1)=p_i$, where we use the fact that $e^{y}>(1+y)$ for $-1<y<0$.
Thereby, the expected value $E[e^{-tX}]$ can be written as
\begin{equation}
\begin{aligned}\label{eq17}
E[e^{-tX}]=\prod_{i=1}^{N}E[e^{-tX_{i}}]<\prod_{i=1}^{N}e^{p_{i}(e^{-t}-1)}=e^{\sum_{i=1}^{N}p_{i}(e^{-t}-1)}=e^{(e^{-t}-1)\mu_{x}}.
\end{aligned}
\end{equation}
Substituting Eq. \eqref{eq17} back into Eq. \eqref{eq16}, the final inequality can be bounded by
\begin{equation}
\begin{aligned}\label{eq18}
{\rm Pr}[X\leq(1-\hat{\delta})\mu_{x}]<\frac{e^{(e^{-t}-1)\mu_{x}}}{e^{-t(1-\hat{\delta})\mu_{x}}}=\left[\frac{e^{-\hat{\delta}}}{(1-\hat{\delta})^{1-\hat{\delta}}}\right]^{\mu_{x}},
\end{aligned}
\end{equation}
where we assume that $t=-\ln(1-\hat{\delta})$ to \textbf{make the bound as tight as possible}. By using Eq.\eqref{eql}, we have
\begin{equation}
\begin{aligned}\label{eqll4}
{\rm Pr}[X\leq(1-\hat{\delta})\mu_{x}]<\left[\frac{e^{-\hat{\delta}}}{(1-\hat{\delta})^{1-\hat{\delta}}}\right]^{\mu_{x}}=g(\mu_{x},-\hat{\delta})=\hat{\epsilon},
\end{aligned}
\end{equation}

Note that the above proof of the multiplicative Chernoff bound exploits the expected value $\mu_{x}$, which means that \textbf{this bound requires the knowledge of $\mu_{x}$}.

\bigskip
\noindent
\textbf{Lemma 3.} A variant of the tight multiplicative Chernoff bound.

Let $X_1, X_2...,X_N$ be a set of independent Bernoulli random variables that satisfy ${\rm Pr}(X_i=1)=p_i$ (not necessarily equal), and let $X:=\sum_{i=1}^NX_i$.
The expected value of $X$ is denoted as $\mu_{x}:=E[X]=\sum_{i=1}^Np_i$. An observed outcome of $X$ is represented as $x$ for a given trial (note that, we have $x\geq0$, $\mu_{x}\geq0$ and $\mu_{x}$ is unknown).
For any $\epsilon>0$, we have that $\mu_{x}$ satisfies
\begin{equation}
\begin{aligned}\label{eq19}
\mu_{x}\geq\underline{\mu_{x}}= \max\{0,x-\Delta(x,\epsilon)\},
\end{aligned}
\end{equation}
with failure probability $\epsilon$, where $\underline{\mu_{x}}$ is the lower bound of $\mu_{x}$ and $\Delta(z,y)$ is the positive root of the following equation
\begin{equation}
\begin{aligned}\label{eq20}
\Delta(z,y)-[z+\Delta(z,y)]\ln\frac{z+\Delta(z,y)}{z}-\ln y=0.
\end{aligned}
\end{equation}
For any $\hat{\epsilon}>0$, we have that $\mu_{x}$ satisfies
\begin{equation}
\begin{aligned}\label{eq21}
\mu_{x}\leq\overline{\mu_{x}}= x+\hat{\Delta}(x,\hat{\epsilon}),
\end{aligned}
\end{equation}
with failure probability $\hat{\epsilon}$, where $\overline{\mu_{x}}$ is upper bound of $\mu_{x}$ and $\hat{\Delta}(z,y)$ is the positive root of the following equation
\begin{equation}
\begin{aligned}\label{eq22}
&\hat{\Delta}(z,y)+z\ln\frac{z}{z+\hat{\Delta}(z,y)}+\ln y=0.\\
\end{aligned}
\end{equation}

\textbf{Proof.}

Here, we first prove the case of Eq. \eqref{eq19}. Obviously, $\underline{\mu_{x}}\equiv0$ if $x\leq\Delta(x,\epsilon)$, otherwise $\underline{\mu_{x}}=x-\Delta(x,\epsilon)$. We consider the case of $x>\Delta(x,\epsilon)$.
We have $x>\underline{\mu_{x}}$ due to $\Delta(x,\epsilon)>0$. The root $\Delta(z,y)$ of Eq. \eqref{eq20} is a monotonic increasing function of $z$ given fixed $y$. The probability can be written as
\begin{equation}
\begin{aligned}\label{eq23}
{\rm Pr}[X\geq\mu_{x}+\Delta(X,\epsilon)]<{\rm Pr}[X>\underline{\mu_{x}}+\Delta(\underline{\mu_{x}},\epsilon)],
\end{aligned}
\end{equation}
where we exploit the fact that the observed outcome $x$ of $X$ for a given trial satisfies $x\geq\underline{\mu_{x}}$, $\mu_{x}\geq\underline{\mu_{x}}$ and $\Delta(z,y)$ is a monotonic increasing function of $z$ given fixed $y$. By using the upper tail of the multiplicative Chernoff bound of \textbf{Lemma 2}, we have
\begin{equation}
\begin{aligned}\label{eq24}
{\rm Pr}[X\geq\underline{\mu_{x}}+\Delta(\underline{\mu_{x}},\epsilon)]<\frac{e^{\Delta(\underline{\mu_{x}},\epsilon)}}{[1+\Delta(\underline{\mu_{x}},\epsilon)/\underline{\mu_{x}}]^{\underline{\mu_{x}}+\Delta(\underline{\mu_{x}},\epsilon)}}.
\end{aligned}
\end{equation}
By using Eq. \eqref{eq20}, we find that
\begin{equation}
\begin{aligned}\label{eq25}
\frac{e^{\Delta(\underline{\mu_{x}},\epsilon)}}{[1+\Delta(\underline{\mu_{x}},\epsilon)/\underline{\mu_{x}}]^{\underline{\mu_{x}}+\Delta(\underline{\mu_{x}},\epsilon)}}=\epsilon.
\end{aligned}
\end{equation}
Therefore, we have the inequality
\begin{equation}
\begin{aligned}\label{eq26}
{\rm Pr}[X\geq\mu_{x}+\Delta(X,\epsilon)]<{\rm Pr}[X>\underline{\mu_{x}}+\Delta(\underline{\mu_{x}},\epsilon)]=\epsilon,
\end{aligned}
\end{equation}
which means that the probability of the observed outcome $x$ of $X$ for a given trial satisfying $x\geq\mu_{x}+\Delta(x,\epsilon)$ is less than $\epsilon$. Combining the results above, we show that $\mu_{x}\geq\underline{\mu_{x}}= \max\{0,x-\Delta(x,\epsilon)\}$ with the failure probability at most $\epsilon$.

Now, we prove the case of Eq. \eqref{eq21}. Obviously, the root $\hat{\Delta}(z,y)$ of Eq. \eqref{eq22} is also a monotonic increasing function of $z$ given fixed $y$. The probability can be written as
\begin{equation}
\begin{aligned}\label{eq27}
{\rm Pr}[X\leq\mu_{x}-\hat{\Delta}(X,\hat{\epsilon})]<{\rm Pr}[X<\overline{\mu_{x}}]={\rm Pr}[X<\overline{\mu_{x}}+\hat{\Delta}(\overline{\mu_{x}},\hat{\epsilon})-\hat{\Delta}(\overline{\mu_{x}},\hat{\epsilon})],
\end{aligned}
\end{equation}
where we exploit the fact that the observed outcome $x$ of $X$ for a given trial satisfies $\hat{\Delta}(x,\hat{\epsilon})>0$ and $\mu_{x} \leq \overline{\mu_{x}}$. By using the lower tail of the multiplicative Chernoff bound of \textbf{Lemma 2}, we have
\begin{equation}
\begin{aligned}\label{eq28}
{\rm Pr}[X<\overline{\mu_{x}}+\hat{\Delta}(\overline{\mu_{x}},\hat{\epsilon})-\hat{\Delta}(\overline{\mu_{x}},\hat{\epsilon})]<\frac{e^{-\hat{\Delta}(\overline{\mu_{x}}
,\hat{\epsilon})}}{\left\{1-\hat{\Delta}(\overline{\mu_{x}},\hat{\epsilon})/[\overline{\mu_{x}}+\hat{\Delta}(\overline{\mu_{x}},\hat{\epsilon})]\right\}^{\overline{\mu_{x}}}}.
\end{aligned}
\end{equation}
By exploiting Eq. \eqref{eq22}, we can find
\begin{equation}
\begin{aligned}\label{eq29}
\frac{e^{-\hat{\Delta}(\overline{\mu_{x}}
,\hat{\epsilon})}}{\left\{1-\hat{\Delta}(\overline{\mu_{x}},\hat{\epsilon})/\left[\overline{\mu_{x}}+\hat{\Delta}(\overline{\mu_{x}},\hat{\epsilon})\right]\right\}^{\overline{\mu_{x}}}}=\hat{\epsilon}.
\end{aligned}
\end{equation}
Therefore, we have the inequality
\begin{equation}
\begin{aligned}\label{eq30}
{\rm Pr}[X\leq\mu_{x}-\hat{\Delta}(X,\hat{\epsilon})]<{\rm Pr}[X<\overline{\mu_{x}}+\hat{\Delta}(\overline{\mu_{x}},\hat{\epsilon})-\hat{\Delta}(\overline{\mu_{x}},\hat{\epsilon})]=\hat{\epsilon},
\end{aligned}
\end{equation}
which means that the probability of the observed outcome $x$ of $X$ for a given trial satisfying $x\leq\mu_{x}-\hat{\Delta}(x,\hat{\epsilon})$ is less than $\hat{\epsilon}$. Combining the results above, we show that $\mu_{x}\leq\overline{\mu_{x}}= x+\hat{\Delta}(x,\hat{\epsilon})$ with the failure probability at most $\hat{\epsilon}$.

Note that the above proof of the variant of the tight multiplicative Chernoff bound does not exploit the expected value $\mu_{x}$, which means that \textbf{this bound does not require the knowledge of $\mu_{x}$}.

\section*{Supplementary Note 3: Comparing with previous methods of statistical fluctuation.}

In this section, we will compare the statistical fluctuation analysis methods proposed in Notes 1 and 2 with previous works.
First, we consider the statistical fluctuation of expected value, given the observed value. Here, we will introduce the rigorous variant of the Chernoff bound method proposed in~\cite{curty2014finite1} and the not-sufficiently-rigorous Gaussian analysis with the central limit theorem.

\bigskip
\noindent
\textbf{Lemma 4.} A variant of the multiplicative Chernoff bound in~\cite{curty2014finite1}.

Let $X_1, X_2, \ldots, X_N$, be a set of independent Bernoulli random variables that
satisfy ${\rm Pr}(X_i=1)=p_i$ (not necessarily equal), and let $X=\sum_{i=1}^N X_i$ and
$\mu_{x}=E[X]=\sum_{i=1}^N p_i$,
where $E[\cdot]$ denotes the mean value. Let
$x$ be
the observed outcome of $X$ for a given trial (i.e., $x\in{\mathbb N}^+$)
and $\mu_{\rm L}=x-\sqrt{N/2\ln{(1/\epsilon)}}$ for
certain $\epsilon>0$. Then,
we have that
$x$ satisfies
\begin{equation}\label{cher_met2}
x=\mu_{x}+\delta,
\end{equation}
except for error probability $\gamma$, where the
parameter $\delta\in[-\Delta,{\hat \Delta}]$.
Let $test_1$, $test_2$ and $test_3$ denote, respectively,
the following three conditions: $\mu_{\rm L}\geq\frac{32}{9}\ln(2\varepsilon^{-1})$,
$\mu_{\rm L}>3\ln(\hat \varepsilon^{-1})$ and $\mu_{\rm L}>\left(\frac{2}{2e-1}\right)^{2}\ln(\hat \varepsilon^{-1})$
for certain $\varepsilon, {\hat \varepsilon}>0$, and
let $g(x,y)=\sqrt{2x\ln{(y^{-1}})}$. Now:
\begin{enumerate}
\item When $test_1$ and $test_2$ are fulfilled, we have that
$\gamma=\epsilon+\varepsilon+{\hat \varepsilon}$,
$\Delta=g(x, \varepsilon^4/16)$ and
${\hat \Delta}=g(x, {\hat \varepsilon}^{3/2})$.
\item When $test_1$ and $test_3$ are fulfilled (and $test_2$ is not fulfilled), we have that
$\gamma=\epsilon+\varepsilon+{\hat \varepsilon}$,
$\Delta=g(x, \varepsilon^4/16)$ and
${\hat \Delta}=g(x, {\hat \varepsilon}^{2})$.
\item When $test_1$ is fulfilled and $test_3$ is not fulfilled, we have that $\gamma=\epsilon+\varepsilon+{\hat \varepsilon}$,
$\Delta=g(x, \varepsilon^4/16)$ and ${\hat \Delta}=\sqrt{(N/2)\ln{(1/\varepsilon)}}$.
\item When
$test_1$ is not fulfilled and $test_2$ is fulfilled, we have that
$\gamma=\epsilon+\varepsilon+{\hat \varepsilon}$,
$\Delta=\sqrt{(N/2)\ln{(1/\varepsilon)}}$ and
${\hat \Delta}=g(x, {\hat \varepsilon}^{3/2})$.
\item When $test_1$ and $test_2$ are not fulfilled, and $test_3$ is fulfilled, we have that
$\gamma=\epsilon+\varepsilon+{\hat \varepsilon}$,
$\Delta=\sqrt{(N/2)\ln{(1/\varepsilon)}}$ and
${\hat \Delta}=g(x, {\hat \varepsilon}^{2})$.
\item When $test_1$, $test_2$ and $test_3$ are not fulfilled, we have that
$\gamma=\varepsilon+{\hat \varepsilon}$,
$\Delta={\hat \Delta}=\sqrt{(N/2)\ln{(1/\varepsilon)}}$.
\end{enumerate}

\begin{figure*}
\centering
\includegraphics[width=10cm]{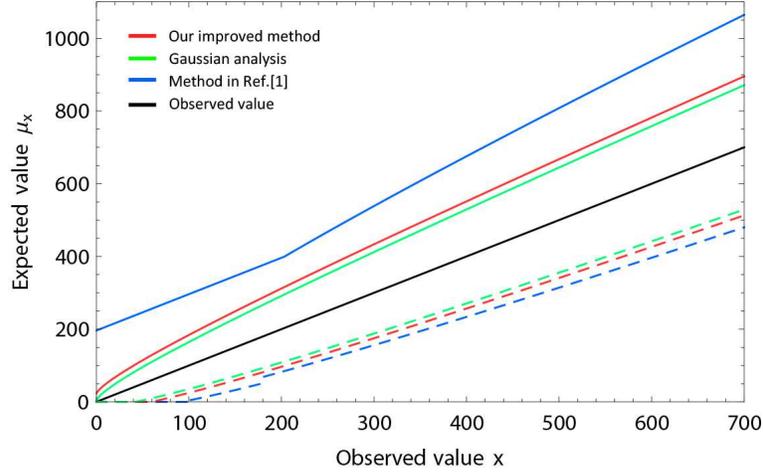}
\caption{Expected value as function of the observed value. The colour solid lines represent the upper bound of the expected value, given the failure probability $\epsilon=10^{-10}$. The colour dotted lines represent the lower bound of the expected value, given the failure probability $\epsilon=10^{-10}$. The black solid line represents the observed value. The results of our improved method are always inferior but comparable to the Gaussian analysis.}\label{s1}
\end{figure*}

\bigskip
\noindent
\textbf{Lemma 5.} Gaussian analysis with the central limit theorem.

Let $X_{1}, X_{2}, \ldots, X_{N}$ be a set of independent and identically distributed Bernoulli random variables that satisfy $\textrm{Pr}(X_{i}=1)=p$, and let $X:=\sum_{i=1}^{N}X_{i}$. The expected value and variance of $X$ are denoted as $\mu_{x}:=E[X]$ and $\sigma^2:=Var[X]$. An observed outcome of $X$ is represented as $x$. When $N\rightarrow\infty$, $\frac{x-\mu_{x}}{\sigma}$ approaches a standard normal distribution $N(0,1)$. Thus, as $N\rightarrow\infty$, $\sigma=\sqrt{x}$, for any fixed $\beta>0$ we have
\begin{equation} \label{central limit theorem}
\begin{aligned}
\textrm{Pr}&[x>\mu_{x}+\beta\sqrt{x}]\rightarrow \frac{1}{\sqrt{2\pi}}\int_{\beta}^{\infty}e^{-\frac{t^2}{2}}dt=\frac{1}{2}\textrm{erfc}(\beta/\sqrt{2}),\\
\textrm{Pr}&[x<\mu_{x}-\beta\sqrt{x}]\rightarrow \frac{1}{\sqrt{2\pi}}\int_{-\infty}^{-\beta}e^{-\frac{t^2}{2}}dt=\frac{1}{2}\textrm{erfc}(\beta/\sqrt{2}),\\
\end{aligned}
\end{equation}
where $\textrm{erfc}(x)=1-\frac{2}{\sqrt\pi}\int_{0}^{x}e^{-t^2}dt$ is the complementary error function.
\newline
\newline

\begin{figure*}
\centering
\includegraphics[width=17cm]{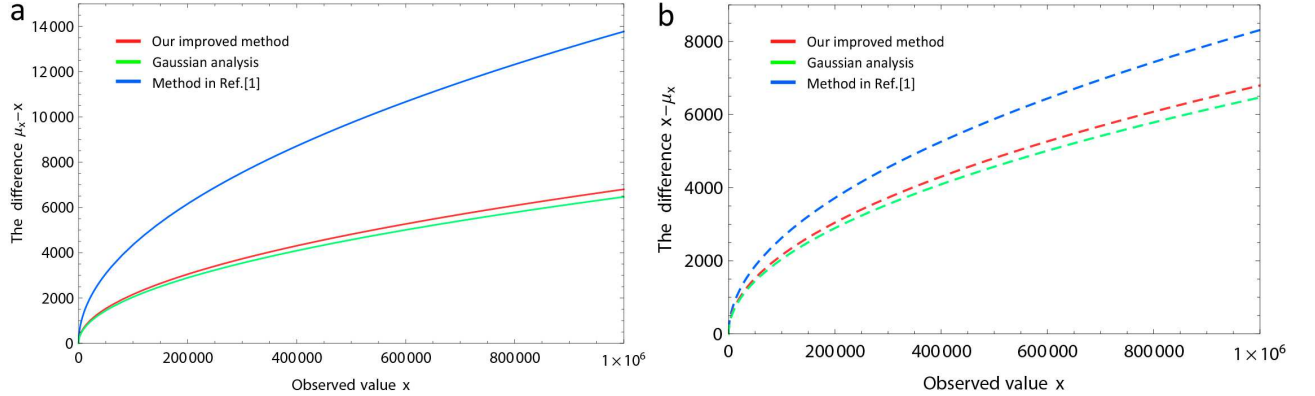}
\caption{The difference between the expected and observed values as function of the observed value. \textbf{a}, The difference between the upper bound of the expected value and observed value. \textbf{ b}, The difference between the observed value and the lower bound of the expected value. The failure probability $\epsilon=10^{-10}$. The results of our improved method are always inferior but comparable to the Gaussian analysis, which means that our rigorous method closes the gap between the rigorous large deviation method in Ref.~\cite{curty2014finite1} and the not-sufficiently-rigorous Gaussian analysis.}\label{s2}
\end{figure*}

The Gaussian analysis requires infinite number of independent and identically distributed Bernoulli random variables. Therefore, any rigorous method with finite number of independent (not necessarily identically distributed) Bernoulli random variables should not be better than Gaussian analysis. Without  loss of generality, we set each failure probability $\epsilon=\hat \epsilon=\varepsilon=\hat\varepsilon=10^{-10}$. Thereby, the three conditions of Lemma 4~\cite{curty2014finite1} become: $test_1$, $\mu_{\rm L}\geq84.33$; $test_2$, $\mu_{\rm L}>69.08$; $test_3$, $\mu_{\rm L}>4.68$. Note that we should have the lower bound $\mu_{x}=x-\Delta\geq\mu_{\rm L}$ in Lemma 4. The three conditions of Lemma 4 further become: $test_1$, $x\geq203$; $test_2$, $x\geq181$; $test_3$, $x\geq102$.
For the quantum key distribution system, the probability ${\rm Pr}(X_i=1)=p_i$ is usually very small, which means $x\ll\sqrt{(N/2)\ln{(1/\epsilon)}}$ and $\Delta=\hat\Delta=\sqrt{(N/2)\ln{(1/\epsilon)}}$ do not apply. Therefore, we can  restate Lemma 4 as: if $x\geq203$, the lower bound of the expected value $\mu_{x}=x-\sqrt{2x\ln(\epsilon^{-3/2})}$ and the upper bound of the expected value $\mu_{x}=x+\sqrt{2x\ln(16\epsilon^{-4})}$; if $181\leq x<203$, the lower bound of the expected value $\mu_{x}=x-\sqrt{2x\ln(\epsilon^{-3/2})}$ and the upper bound of the expected value $\mu_{x}=x+\sqrt{2\times203\ln(16\epsilon^{-4})}$; if $102\leq x<181$, the lower bound of the expected value $\mu_{x}=x-\sqrt{2x\ln(\epsilon^{-2})}$ and the upper bound of the expected value $\mu_{x}=x+\sqrt{2\times203\ln(16\epsilon^{-4})}$; if $x<102$, the lower bound of the expected value $\mu_{x}=0$ and the upper bound of the expected value $\mu_{x}=x+\sqrt{2\times203\ln(16\epsilon^{-4})}$. Note that $\epsilon=10^{-10}$ and we exploit the fact that $\Delta$ is the monotonic increasing function of $x$ given fixed failure probability $\epsilon$.

Figures \ref{s1} and \ref{s2} compare the results among our improved method, the large deviation method in Ref.~\cite{curty2014finite1}, and the Gaussian analysis. The lower bound of the expected value in Gaussian analysis is always $\mu_{x}=0$, given the observed value $x\leq41$. The upper bound of the expected value in Gaussian analysis is $\mu_{x}=0$, given the observed value $x=0$. The lower bound of the expected value in our improved method is always $\mu_{x}=0$, given the observed value $x\leq59$. The upper bound of the expected value in our improved method is $\mu_{x}=\ln\epsilon^{-1}=23.0259$, given the observed value $x=0$. The lower bound of the expected value in the large deviation method in Ref.~\cite{curty2014finite1} is always $\mu_{x}=0$, given the observed value $x\leq101$. The upper bound of the expected value in the large deviation method in Ref.~\cite{curty2014finite1} is $\mu_{x}=\sqrt{406\ln(16\epsilon^{-4})}=196.264$, given the observed value $x=0$.
The results of our improved method are always inferior but comparable to the Gaussian analysis, which means that our rigorous method closes the gap between the rigorous large deviation method in Ref.~\cite{curty2014finite1} and the not-sufficiently-rigorous Gaussian analysis.

\begin{figure*}
\centering
\includegraphics[width=17cm]{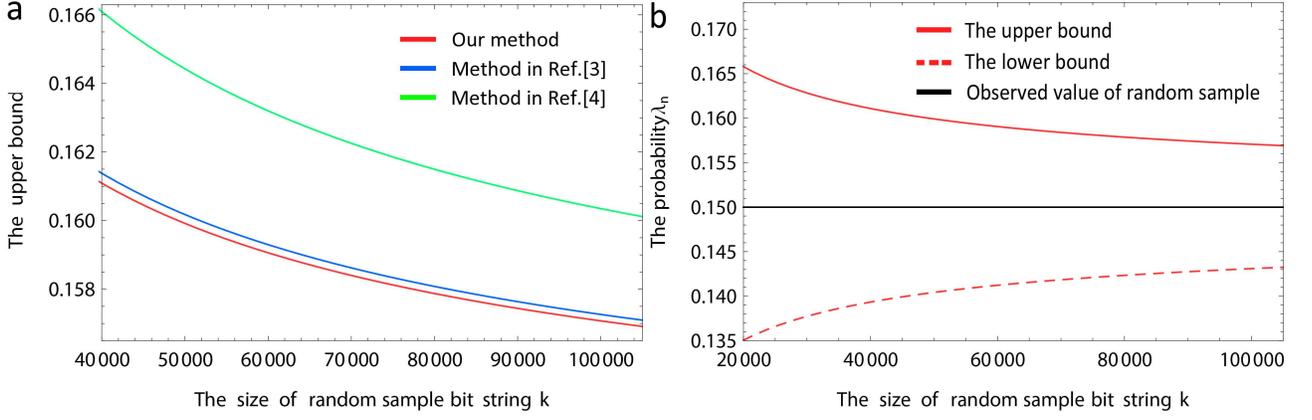}
\caption{Comparing the methods of random sampling without replacement. \textbf{a}, The upper bound probability of bit value 1 observed in remaining bit string, given $n=10^{6}$, $\lambda_{k}=0.15$ and $\epsilon=10^{-10}$. \textbf{b}, The upper and lower bound probabilities of bit value 1 observed in remaining bit string in our improved method, given $n=10^{6}$, $\lambda_{k}=0.15$ and $\epsilon=10^{-10}$.}\label{s3}
\end{figure*}

Second, we consider the statistical fluctuation of random sampling without replacement. The problem of random sampling without replacement is usually solved by the Serfling inequality~\cite{tomamichel2012tight1}. However, the Serfling inequality cannot give very good bound here since this result does not consider the properties of the priori distribution. By using the hypergeometric function distribution, one can provide a good bound even in a high-loss regime~\cite{Fung:2010:Practical1,lim2014concise1}.

\bigskip
\noindent
\textbf{Lemma 6.} The upper bound tail inequality of random sampling without replacement~\cite{Fung:2010:Practical1,lim2014concise1}.

Let ${\mathcal X_{n+k}}:=\{x_{1},x_{2},\cdots,x_{n+k}\}$ be a string of binary bits with $n+k$ size, in which the number of bit value 1 is unknown. Let ${\mathcal X_{k}}$ be a random sample (without replacement) bit string with $k$ size from ${\mathcal X_{n+k}}$. Let $\lambda_{k}$ be the probability of bit value 1 observed in ${\mathcal X_{k}}$. Let ${\mathcal X_{n}}$ be the remaining bit string, where the probability of bit value 1 observed in ${\mathcal X_{n}}$ is $\lambda_{n}$.
For any $\epsilon>0$, we have the upper tail
\begin{equation}
\begin{aligned}\label{}
{\rm Pr}[\lambda_{n}\geq \lambda_{k}+\gamma(n,k,\lambda_{k},\epsilon)]\leq\epsilon,
\end{aligned}
\end{equation}
where $\gamma(a,b,c,d)$ is the positive root of the following equation~\cite{Fung:2010:Practical1}
\begin{equation}
\begin{aligned}\label{eq38}
h\left[c+\frac{a}{a+b}\gamma(a,b,c,d)\right]-\frac{b}{a+b}h[c]-\frac{a}{a+b}h[c+\gamma(a,b,c,d)]-\frac{1}{2(a+b)}\log_{2}\frac{a+b}{abc(1-c)d^{2}}=0,
\end{aligned}
\end{equation}
where $h[x]=-x\log_{2}x-(1-x)\log_{2}(1-x)$ is the Shannon entropy function. By exploiting the Taylor expansion, the above result can be written as an approximate analytical formula~\cite{lim2014concise1},
\begin{equation}
\begin{aligned}\label{eq39}
\gamma(a,b,c,d)]=\sqrt{\frac{(a+b)c(1-c)}{ab\ln2}\log_{2}\frac{a+b}{abc(1-c)d^{2}}}.
\end{aligned}
\end{equation}
Note that the approximate analytical formula is only true for appropriate parameters $a$ and $b$, which means that the result of approximate analytical formula Eq. \eqref{eq39} is larger than Eq. \eqref{eq38}. The approximate analytical formula is not true, given small $a$ and $b$.

Figure \ref{s3} compares the results among our improved method, the methods in Ref.~\cite{Fung:2010:Practical1} and Ref.~\cite{lim2014concise1} for the random sampling without replacement. The probability of bit vale 1 observed in random sample bit string is $\lambda_{k}=0.15$. The size of remaining bit strings is $n=10^{6}$. Our bound is the tightest because we avoid excessive inequality scaling. Furthermore, we provide the lower bound tail inequality for random sampling without replacement, which is shown in Fig. \ref{s3}b.

\section*{Supplementary Note 4: Decoy-state analysis with three-intensity}
Here, we exploit the decoy-state method with three-intensity ($0<\omega<\nu$)~\cite{grasselli2019practical1,Yin:2018:Twin1} to estimate the upper bound of the expected yield $\overline{Y}_{n,m}^{*}$. The upper bound of the expected yield $\overline{Y}_{0,0}^{*}=\overline{Q}_{0,0}^{*}$. The upper bound of the expected yield $\overline{Y}_{1,1}^{*}$, $\overline{Y}_{0,2}^{*}$ and $\overline{Y}_{2,0}^{*}$ can be given by
\begin{equation}
\begin{aligned}\label{}
\overline{Y}_{1,1}^{*}=\frac{e^{2\omega}\overline{Q}_{\omega,\omega}^{*}-e^{\omega}(\underline{Q}_{\omega,0}^{*}+\underline{Q}_{0,\omega}^{*})+\overline{Q}_{0,0}^{*}}{\omega^{2}},\\
\overline{Y}_{0,2}^{*}=\frac{\omega e^{\nu}\overline{Q}_{0,\nu}^{*}-\nu e^{\omega}\underline{Q}_{0,\omega}^{*}+(\nu-\omega)\overline{Q}_{0,0}^{*}}{\nu\omega(\nu-\omega)/2},\\
\overline{Y}_{2,0}^{*}=\frac{\omega e^{\nu}\overline{Q}_{\nu,0}^{*}-\nu e^{\omega}\underline{Q}_{\omega,0}^{*}+(\nu-\omega)\overline{Q}_{0,0}^{*}}{\nu\omega(\nu-\omega)/2}.\\
\end{aligned}
\end{equation}
The upper bound of the expected yield $\overline{Y}_{0,4}^{*}$ and $\overline{Y}_{4,0}^{*}$ can be given by
\begin{equation}
\begin{aligned}\label{}
\overline{Y}_{0,4}^{*}= {\rm min}\left\{1,~\frac{\omega e^{\nu}\overline{Q}_{0,\nu}^{*}-\nu e^{\omega}\underline{Q}_{0,\omega}^{*}+(\nu-\omega)\overline{Q}_{0,0}^{*}}{\nu\omega(\nu^{3}-\omega^{3})/4!}\right\},\\
\overline{Y}_{4,0}^{*}= {\rm min}\left\{1,~\frac{\omega e^{\nu}\overline{Q}_{\nu,0}^{*}-\nu e^{\omega}\underline{Q}_{\omega,0}^{*}+(\nu-\omega)\overline{Q}_{0,0}^{*}}{\nu\omega(\nu^{3}-\omega^{3})/4!}\right\}.\\
\end{aligned}
\end{equation}
The upper bound of the expected yield $\overline{Y}_{1,3}^{*}$ and $\overline{Y}_{3,1}^{*}$ can be given by
\begin{equation}
\begin{aligned}\label{}
\overline{Y}_{1,3}^{*}
={\rm min}\left\{1,\frac{(\omega e^{\nu+\omega}\overline{Q}_{\omega,\nu}^{*}+(\nu-\omega)e^{\omega}\overline{Q}_{\omega,0}^{*}+\nu e^{\omega}\overline{Q}_{0,\omega}^{*})-(\omega e^{\nu}\underline{Q}_{0,\nu}^{*}+\nu e^{2\omega}\underline{Q}_{\omega,\omega}^{*}+(\nu-\omega)\underline{Q}_{0,0}^{*})}{\nu\omega^{2}(\nu^{2}-\omega^{2})/3!}\right\}\\
\overline{Y}_{3,1}^{*}
={\rm min}\left\{1,\frac{(\omega e^{\nu+\omega}\overline{Q}_{\nu,\omega}^{*}+\nu e^{\omega}\overline{Q}_{\omega,0}^{*}+(\nu-\omega) e^{\omega}\overline{Q}_{0,\omega}^{*})-(\omega e^{\nu}\underline{Q}_{\nu,0}^{*}+\nu e^{2\omega}\underline{Q}_{\omega,\omega}^{*}+(\nu-\omega)\underline{Q}_{0,0}^{*})}{\nu\omega^{2}(\nu^{2}-\omega^{2})/3!}\right\}\\
\end{aligned}
\end{equation}
The upper bound of the expected yield $\overline{Y}_{2,2}^{*}$ can be given by
\begin{equation}
\begin{aligned}\label{}
\overline{Y}_{2,2}^{*}
={\rm min}\Big\{1,\frac{1}{\nu^{2}\omega^{2}(\nu-\omega)^{2}/4}[\omega^{2}e^{2\nu}\overline{Q}_{\nu,\nu}^{*}+\nu^{2}e^{2\omega}\overline{Q}_{\omega,\omega}^{*}+\omega(\nu-\omega) e^{\nu}(\overline{Q}_{\nu,0}^{*}+\overline{Q}_{0,\nu}^{*})+(\nu-\omega)^{2}\overline{Q}_{0,0}^{*}]\\
-[\nu\omega e^{\nu+\omega}(\underline{Q}_{\nu,\omega}^{*}+\underline{Q}_{\omega,\nu}^{*})+\nu(\nu-\omega)e^{\omega}(\underline{Q}_{\omega,0}^{*}+\underline{Q}_{0,\omega}^{*})]\Big\}.\\
\end{aligned}
\end{equation}
Let $\overline{s}_{n,m}^{*}=\overline{Y}_{n,m}^{*}N_{\rm X}\sum_{a,b}p_{a}p_{b}P_{n}^{a}P_{m}^{b}$ be upper bound of the expected bit  number in the ${\rm X}$ basis
given that Alice and Bob send $n$-photon and $m$-photon. By using the upper tail of the multiplicative Chernoff bound in Lemma 2, we can estimate the upper bound of
the observed bit number $\overline{s}_{n,m}$ given by $\overline{s}_{n,m}^{*}$  with failure probability $\epsilon_{2}$.  Thereby, the upper bound of the observed yield
$\overline{Y}_{n,m}=\overline{s}_{n,m}/\left(N_{\rm X}\sum_{a,b}p_{a}p_{b}p_{n}^{a}p_{m}^{b}\right)$. For the case of $n+m\geq5$, we let the upper bound of the observed yield $\overline{Y}_{n,m}=1$.

In our simulation, we have~\cite{Yin:2018:Twin1}
\begin{equation}
\begin{aligned}\label{}
Q_{\rm Z}&=(1-p_{d})[1-(1-2p_{d})e^{-2\mu\eta}],\\
E_{\rm Z}&=[e_{d_{\rm Z}}Q_{\rm Z}^{C}+(1-e_{d_{\rm Z}})Q_{\rm Z}^{E}]/Q_{\rm Z},\\
Q_{\rm Z}^{E}&=p_{d}(1-p_{d})e^{-2\mu\eta},\\
Q_{\rm Z}^{C}&=(1-p_{d})[1-(1-p_{d})e^{-2\mu\eta}],\\
Q_{a,b}&=2(1-p_{d})e^{-\frac{1}{2}(a+b)\eta}I_{0}(\sqrt{ab}\eta)-2(1-p_{d})^{2}e^{-(a+b)\eta}.\\
\end{aligned}
\end{equation}
where $I_{0}(x)$ is the modified Bessel function of the first kind, $p_{d}$ is the dark count rate, $e_{d_{\rm Z}}$ is the misalignment rate of the ${\rm Z}$ basis,
$\eta$ is the overall efficiency between Alice (Bob) and Charlie.



\end{document}